\PassOptionsToPackage{left=2cm,right=2cm,top=2cm,bottom=2cm}{geometry}

\documentclass[sn-vancouver-num]{sn-jnl}

\usepackage{graphicx}%
\usepackage{multirow}%
\usepackage{amsmath,amssymb,amsfonts}%
\usepackage{amsthm}%
\usepackage{mathrsfs}%
\usepackage[title]{appendix}%
\usepackage{xcolor}%
\usepackage{textcomp}%
\usepackage{manyfoot}%
\usepackage{booktabs}%
\usepackage{algorithm}%
\usepackage{algorithmicx}%
\usepackage{algpseudocode}%
\usepackage{listings}%
\usepackage[none]{hyphenat}
\usepackage{placeins}
\usepackage{url}
\usepackage{multirow}
\usepackage[capitalize]{cleveref}
\usepackage{tabularx}
\usepackage{tikz}

\theoremstyle{thmstyleone}%

\theoremstyle{thmstyletwo}%
\theoremstyle{thmstylethree}%

\usepackage{microtype}                
\usepackage{url}                      %
\makeatletter
\@twosidefalse
\@mparswitchfalse
\makeatother

\begin{document}

\title[Article Title]{Rank-Constrained Deep Matrix Completion for Group Recommendation}
 
\author*[1]{\fnm{Mubaraka } \sur{Sani Ibrahim}}\email{msani@aust.edu.ng}

\author[2]{\fnm{Lehel} \sur{Csat\'o}}\email{lehel.csato@ubbcluj.ro}

\author[3]{\fnm{Isah Charles} \sur{Saidu}}\email{charles.isah@bazeuniversity.edu.ng}
\affil*[1]{\orgdiv{Department of Computer Science}, \orgname{African University of Science and Technology}, \orgaddress{\state{Abuja},  \country{Nigeria}}}

\affil[2]{\orgdiv{Faculty of Mathematics and Computer Science}, \orgname{Babes-Bolyai University}, \state{Cluj-Napoca}, \country{Romania}}

\affil[3]{\orgdiv{Department of Computer Science}, \orgname{Baze University}, \state{Abuja},  \country{Nigeria}}

\abstract{The growing popularity of group activities has increased the need for methods that provide recommendations to groups of users given their individual preferences. Many existing group recommender systems rely on aggregating individual user preferences, but they often struggle with high-dimensional and highly sparse rating data commonly found in real-world scenarios. We propose Group Rank-Constrained Deep Matrix Completion (Group RC-DMC), a novel framework that extends RC-DMC by integrating group-level representation learning via a Set-Transformer aggregator, jointly leveraging low-rank structure and attention-based non-linear modeling. Unlike most existing group recommender systems, Group RC-DMC unifies explicit low-rank regularization, linear encoder–decoder architectures, and attention-based non-linear group modelling within a single framework, yielding accurate predictions at both the individual and group levels. Group RC-DMC addresses data sparsity through low-rank matrix completion, computing per-user latent representations from observed ratings only, and enforcing a rank constraint on the latent space using a nuclear-norm proximal step (periodic singular value thresholding). The decoder is parametrised as a low-rank factorization, enabling efficient inference. Experimental results on the \textit{MovieLens} and \textit{Goodbooks} datasets demonstrate that Group RC-DMC achieves superior reconstruction accuracy (lower group RMSE), while remaining computationally efficient and competitive group-level performance in terms of precision, recall, and F1 score compared to weighted-before-factorization (WBF) and after-factorization (AF) baselines. The results highlight the model’s ability to recover the underlying low-rank structure of user-item interactions and provide robust group recommendations across small, medium, and large user groups.}


\maketitle 
\raggedbottom

\section{Introduction}

A recommender system is an algorithm that offers personalized recommendations to users based on preferences, past behaviour, and similarities to other users. Although early applications focused on movies and entertainment, recommender systems are now widely adopted across various domains, including music, television, e-commerce, tourism, and social platforms, where they play a vital role in shaping the user experience.  
In these systems, users typically express their preferences for items by providing ratings. However, not all items receive ratings from users, resulting in a partially observed, high-dimensional user-item matrix with missing entries. This challenge is commonly addressed through \emph{matrix completion}, which aims to recover missing entries in the rating matrix \cite{CandesTao2009}.  

A core assumption underlying matrix completion is that user-item matrices can be represented in a low-dimensional latent space. Classical approaches, such as matrix factorization and regularization techniques, like nuclear norm minimization, exploit this low-rank structure to approximate the rating matrix. By imposing a low-rank constraint on the solution, these methods reduce sample complexity and allow algorithms to efficiently scale to large datasets \cite{CandesTao2009,Keshavan2010}.  
Despite these strengths, recommender systems face two persistent challenges: the high sparsity of observations, as users tend to rate only a small portion of items \cite{Keshavan2010}, and the presence of complex, non-linear interactions shaped by contextual and latent factors, which simple linear models struggle to capture effectively \cite{He2017}.

A group recommender system generates recommendations for a collection of users by aggregating their individual preferences into a shared recommendation that satisfies the group as a whole \cite{jameson2007}. Group recommender systems differ from traditional recommenders by considering collective preferences, interactions, and potential conflicts among group members. This involves balancing diverse interests, ensuring equity, and modelling group dynamics \cite{Li2024survey} through various strategies, including weighted aggregation, majority voting, or consensus-based rules. 

Practical group recommendations depend crucially on how groups are defined. In this work, we assume 
that group memberships are provided as part of the input data and focus on modelling group-level preferences given these predefined groups. In this work
we focus on modelling the \emph{group} as a set of individual users and on learning a pooled,
permutation-invariant representation of the set; the exact procedures used to form groups for
evaluation are described in Section~\ref{sec:methodology} so that experimental results are reproducible and
interpretable.

Existing group recommendation methods often rely on aggregation strategies to combine the preferences of individual members into a single group profile. Among these, matrix factorization based approaches are widely used \cite{Ortega2016}.
Although these methods are straightforward and interpretable, they typically overlook non-linear interaction patterns within groups and impose no principled structural constraints on group-level embeddings or decoders.

Consequently, few deep learning models are designed to capture both the intricate dynamics of group interactions and impose explicit low‑rank constraints, creating a methodological gap for neural approaches that retain the advantages of classical low‑rank recovery.

Models such as AutoEncoder based matrix completion (AEMC) \cite{FanChow2017}, it's deep variants (DLMC) \cite{FanChow2017}, and neural collaborative filtering \cite{He2017} learn highly expressive latent embeddings, but typically rely on unconstrained neural mappings to encode and reconstruct user-item interactions. As a result, they tend to overfit under high sparsity and exhibit poor generalization \cite{Nguyen2018}.

While our experimental evaluation focuses on MovieLens and Goodbooks to allow controlled comparison with prior group recommender work, it is worth noting that substantially larger-scale datasets such as the Netflix Prize dataset with over 100 million ratings from approximately 480,000 users on 17,770 movies~\cite{netflixprize}, the Amazon Reviews dataset containing hundreds of millions of product reviews~\cite{amazonreviews}, and the Yelp Open Dataset with millions of user reviews~\cite{yelpopen} also exist, introducing additional challenges related to scale and computational resources.

The remainder of this paper is organized as follows. 
Section~2 reviews related work, covering matrix completion (Section~2.1), autoencoder-based matrix completion and deep variants (Section~2.2) and set transformers for permutation-invariant representation (Section~2.3).  
Section~3 introduces the proposed Group rank-constrained deep matrix completion methodology, including the integration of Set transformers for modelling group interactions (Section~3.1). 
Section~4 presents the experimental evaluation and comparative results. 
Finally, Section~5 concludes the paper with a summary of findings and directions for future research.
\section{Related Work and Background }
This section reviews the foundations of matrix completion and low-rank modelling, surveys neural extensions for collaborative filtering, and discusses group recommendation strategies. We also highlight the role of permutation-invariant architectures such as Set transformers in learning group representations. Together, these strands of research motivate our proposed RC-DMC framework.

\subsection{Notation}
We provide a brief summary of notations used throughout the paper; we adopt notation similar to \cite{Cai2010,Mazumder2010}. 
Let $X \in \mathbb{R}^{m \times n}$ denote the partially observed user–item rating matrix, 
where $m$ is the number of items and $n$ is the number of users. Define a matrix $P_{\Omega}(X)$ 
 \[
P_\Omega(X) =
\begin{cases}
X_{i,j} \quad if \, (i,j)  \in  \Omega \\
0 \qquad if \, (i,j)  \not\in \Omega
\end{cases}
\]
 as the projection of the matrix $X_{m \times n}$ onto the observed entries. 
Here, $\Omega \subset \{1,\ldots,m\} \times \{1,\ldots,n\}$ denotes the index set of observed entries. 
For $(i,j) \in \Omega$, the observed rating is $X_{ij}$, while $X_{ij}=0$ for unobserved entries. 
Following \cite{Mazumder2010}, the complementary projection $P_{\Omega^\perp}(X)$ is defined on the unobserved entries such that
\[
P_{\Omega^\perp}(X) + P_\Omega(X) = X,
\]
where $\Omega^\perp$ denotes the complement of $\Omega$.
In our setting, the rating takes values in the  range $[1-5]$, and the value 0 is reserved exclusively to denote missing (unobserved) entry.

\subsection{Matrix Completion}

Matrix completion aims to recover the missing entries of a partially observed matrix from its observed entries, with the objective of reconstructing a close approximation to the underlying full matrix. Obviously, this is an ill-posed problem and usually requires additional assumptions to the problem. To obtain a well-posed problem, it is necessary to make low-rank assumptions about the underlying matrix \cite{CandesPlan2009}. Netflix’s user–movie rating matrix is a classic example of data believed to be low rank, as users’ movie preferences are influenced by a small number of latent factors, such as movie genre and audience type. This assumption is grounded in the empirical observation that user preferences and item attributes are often governed by a small set of latent factors, which can be captured effectively by low-dimensional embeddings \cite{KorenBellVolinsky2009}. Low-rank techniques are effective in addressing matrix completion problems by exploiting redundant information in interaction matrices and compressing it for improved computational performance. 
There are two classical approaches to matrix completion: low-rank matrix factorization, which reformulates the problem as an optimization task that learns two (or more) low-dimensional factor matrices, and nuclear norm minimization, which formulates matrix recovery as a convex optimization problem that penalizes the sum of singular values to promote low-rank solutions \cite{Emmanuel2009}. 

Several approaches in the literature have been proposed to address the high‑dimensional and highly sparse nature of recommender system data. Another widely used approach is Alternating Least Squares (ALS), which factorizes the user-item rating matrix by alternately fixing one factor matrix (users or items) while solving for the other, thereby minimizing the squared error between observed and predicted ratings. ALS is particularly attractive for large-scale recommender systems because its updates parallelize efficiently, allowing the algorithm to scale to massive datasets with millions of users and items \cite{Hu2008}.

Nuclear norm minimization is a powerful convex relaxation technique for recovering and completing low-rank matrices. The central idea involves replacing the non-convex rank function with the nuclear norm, thereby converting a difficult rank minimization problem into a convex optimization problem. Research on this method includes \cite{Emmanuel2009}, which addresses the challenge of recovering a low‑rank matrix from only a subset of observed entries, a problem central to collaborative filtering and recommendation systems. They showed that, under certain conditions, most low‑rank matrices can be reconstructed by solving a convex optimization problem that minimizes the nuclear norm subject to the observed entries. 

Another study introduces Soft‑Impute algorithm, a convex optimization method that applies iterative soft‑thresholding of singular values to effectively address large‑scale matrix completion problems \cite{Mazumder2010}.

Classical low‑rank approaches, such as matrix factorization and nuclear norm minimization, exploit the hypothesis that user-item preferences lie in a
low-dimensional subspace, which both reduces sample complexity and yields scalable
algorithms for very large datasets.

\subsection{Autoencoder-based matrix completion and deep variants}
To address non-linearities that classical low-rank models cannot capture, the recommender community has increasingly adopted deep neural architectures.
In this section, we review the foundational matrix completion techniques that motivate our proposed method. We begin with AutoEncoder-based Matrix Completion (AEMC), a
non-linear alternative to traditional collaborative filtering approaches. We then discuss its extension to deep architectures.

AEMC is a neural network approach designed to estimate missing entries in a partially observed matrix by reconstructing observed patterns through low‑dimensional embeddings \cite{FanChow2017}. Given an incomplete user-item matrix \( X \in \mathbb{R}^{m \times n} \), AEMC jointly learns an encoder–decoder architecture and imputes missing values by minimizing the reconstruction loss over the observed entries.
The optimization problem is typically formulated as:

\begin{equation}\label{eq1}
\min_{\theta, X} \frac{1}{2n} \sum_{i=1}^{n} \left\| \Omega_i \odot \left( x_i - g^{(2)}(W^{(2)} g^{(1)}(W^{(1)} x_i + b^{(1)}) + b^{(2)}) \right) \right\|_2^2 + \frac{\lambda}{2} \left( \|W^{(1)}\|_F^2 + \|W^{(2)}\|_F^2 \right)
\end{equation}

Here, \( \theta = \{W^{(1)}, b^{(1)}, W^{(2)}, b^{(2)}\} \) are the learnable parameters of the autoencoder, \( g^{(1)} \) and \( g^{(2)} \) are element-wise activation functions (e.g., ReLU, sigmoid), and \(\Omega_i \in \{0,1\}^n \) denotes  a binary mask indicating observed entries in the input vector \( x_i \). Reconstruction loss is regularized via the Frobenius norms of the weight matrices to prevent overfitting.

DLMC extends the AEMC approach by using deeper, stacked autoencoder architectures to better capture complex non-linear mappings between observed and latent variables. The deeper structure allows the model to learn hierarchical feature representations and improves generalization, especially in highly sparse settings.
The training objective for a DLMC model with \( k \) layers is:

\begin{equation}\label{eq2}
\min_{\theta, X} \frac{1}{2n} \sum_{i=1}^{n} \left\| \Omega_i \odot \left( x_i - g^{(k)}(g^{(k-1)}(\dots g^{(1)}(x_i, \theta^{(1)}) \dots, \theta^{(k-1)}), \theta^{(k)}) \right) \right\|_2^2 + \frac{\lambda}{2} \sum_{j=1}^{k} \|W^{(j)}\|_F^2
\end{equation}
where \( \theta^{(j)} = \{W^{(j)}, b^{(j)}\} \) represents the parameters of the \( j \)-th layer, and \( g^{(j)} \) is the activation function at that layer. DLMC models are commonly trained using greedy layer-wise pre-training, followed by global fine-tuning \cite{Bengio2006}.
Directly solving problem equation ~\eqref{eq2} is challenging due to its susceptibility to poor local minima. To mitigate this, the problem can be reformulated using stacked autoencoders and optimized via greedy layer-wise pre-training, as suggested in \cite{Bengio2006}.

\subsection{Set Transformers for Permutation-Invariant Representation}

In group recommendation tasks, where a group is represented as a set of individual user embeddings, it is crucial to model the set without relying on any particular order. Traditional neural architectures are not inherently permutation-invariant. Set transformers \cite{Lee2019SetTransformer} address this by leveraging self-attention mechanisms tailored to unordered input.

Set Transformers are composed of \textit{Set Attention Blocks (SAB)} and \textit{Pooling by Multihead Attention (PMA)}.
SAB modules leverage multi-head self-attention to model pairwise interactions and higher-order interactions among elements within the input set. PMA layers then aggregate the resulting representations into a fixed-size summary vector by attending to the set via a set of learnable seed vectors.
    
This architecture makes Set transformers especially well-suited for learning robust group-level representations in recommender systems.

\subsection{Problem Statement: Group Recommendation}\label{sec:group_recommendation}
Let $\mathcal{U} = \{u_i\}_{i=1}^m$ denote the set of $m$ users and 
$\mathcal{V} = \{v_j\}_{j=1}^n$ the set of $n$ items. 
The interaction between users and items is represented by the partially observed rating matrix 
$X \in \mathbb{R}^{m \times n}$, where the $i$-th row $x_i$ corresponds to the ratings of user $u_i$ across all items. 

Given a group of users $G \subset \mathcal{U}$, the task is to recommend a list of $k$ items, 
denoted by $V_G = \{v_j\}_{j=1}^k$ with $V_G \subset \mathcal{V}$, 
that best satisfy the collective preferences of the group $G$.

\section{Methodology}\label{sec:methodology}
In this section, we present the methodology underlying our proposed approach for group recommendation. We begin by introducing the key components of our model, including low-rank initialization,  encoder-decoder structures, and attention-based group aggregation. We then analyse the computational complexity and convergence behaviour of the method, and finally describe the baseline techniques used for comparison.

\subsection{Group Rank-Constrained Deep Matrix Completion}

\begin{figure}[t]
  \centering
  \includegraphics[width=0.9\textwidth]{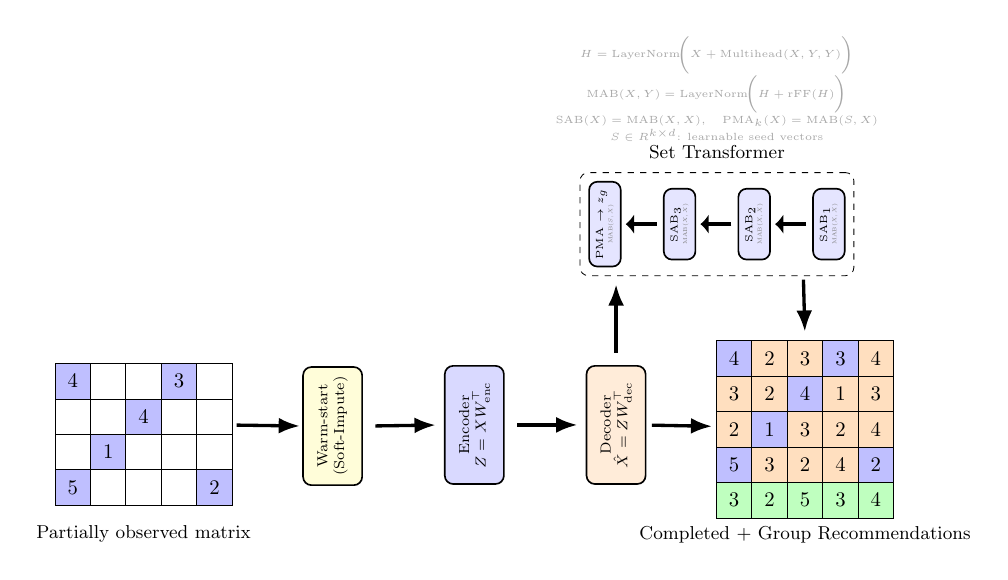}
  \caption{%
Group RC-DMC begins with warm-start initialization using Group Soft-Impute, producing low-rank latent embeddings from partially observed ratings. 
Individual preferences are encoded and reconstructed through a linear encoder–decoder with an explicit low-rank constraint. 
Group dynamics are then modelled using a Set Transformer composed of stacked Set Attention Blocks (SAB) and Pooling by Multihead Attention (PMA), capturing non-linear interactions among group members and aggregating them into unified, permutation-invariant group representations for personalised group recommendation.
}
  \label{fig1}
\end{figure}

We  propose \emph{Group rank-constrained deep matrix completion} (Group RC-DMC), a framework that combines expressive neural encoders and decoders with explicit low-rank regularization, as illustrated in Fig.~\ref{fig1}. For group recommendation, RC-DMC is enhanced with a Set transformer module, where stacked self-attention blocks and pooling-by-multihead-attention capture non-linear interactions among group members, yielding a unified group representation within the same low-rank structure and further detailed in Algorithm~\ref{algo2} and Algorithm~\ref{algo3}.

The approach begins with a warm-start initialization using Group soft-impute \cite{Ibrahim2025} as shown in Algorithm ~\ref{algo1}. 
For each group \(G\subseteq\{1,\dots,m\}\) we define the group-average rating row and an item weight used to form the augmented matrix. Specifically,
\begin{align}
\label{eq:group_avg}
r_{G,j} & = \frac{1}{|G|}\sum_{i\in G} x_{i,j}, \qquad j\in\Omega,\\
\label{eq:group_weight}
w_{G,j} & = \frac{\#\{i\in G \mid x_{i,j}\ \text{observed}\}}{|G|}\cdot \frac{1}{1+\sigma_{G,j}},
\end{align}
where \(\sigma_{G,j}\) is the standard deviation of the observed ratings for item \(j\) within group \(G\). The weighted group-average row is \(w_G\odot r_G\) (elementwise product). 
We then form an augmented matrix \(X_{\mathrm{aug}}\) that appends weighted group-average rows to the observed data, as defined in Eq.~\eqref{eq:xaug}, and run iterative soft-threshold SVD on the observed entries.

\begin{equation}
\label{eq:xaug}
X_{\mathrm{aug}} \;=\; \begin{bmatrix} X \\[3pt] w_G\odot R_G \end{bmatrix},
\end{equation}

Starting from an initial estimate \(Z^{(0)}=0\), we apply an iterative Soft-Impute procedure in which, for a decreasing sequence of thresholds from \(\tau_{\max}\) to \(\tau_{\min}\), we perform singular value soft-thresholding.
At each iteration, we compute
\begin{equation}
U, D, V^\top \gets \mathrm{SVD}\big(P_\Omega(X_{\mathrm{aug}}) + P_{\Omega^c}(Z^{(old)})\big),
\end{equation}
apply singular value shrinkage \(D_\tau = \max(D - \tau, 0)\), and update the low-rank estimate as
\begin{equation}\label{eq:svt_def}
Z^{(t)} = U D_\tau V^\top.
\end{equation}

The iterations terminate when the relative change
\(\|Z^{(new)}-Z^{(old)}\|_F / \|Z^{(old)}\|_F\) falls below a tolerance \(\epsilon\).
Each Soft-Impute iteration corresponds to solving the following nuclear-norm regularized problem:
\begin{equation}\label{eqn3}
\min_{Z_{\mathrm{SI}}} \; \|P_{\Omega}(X_{\mathrm{aug}}) - P_{\Omega}(Z_{\mathrm{SI}})\|_F^2 + \tau \|Z_{\mathrm{SI}}\|_*,
\end{equation}
where \(\|\cdot\|_F\) denotes the Frobenius norm, \(\|\cdot\|_*\) denotes the nuclear norm, and \(\tau>0\) is the singular-value threshold.
The converged solution \(Z_{\mathrm{SI}}\) is used to initialize the encoder weights \(W_{\mathrm{enc}}\) via a least-squares fit by solving
\begin{equation}\label{eqn4}
W_{\mathrm{enc}} = \arg\min_W \|X_{\mathrm{aug}} W^\top - Z_{\mathrm{SI}}\|_F^2,
\end{equation}
ensuring the encoder is aligned with a low-rank structure from the outset. 

Next, each user rating vector \(x_i \in \mathbb{R}^n\) is mapped to a \(d\)-dimensional latent code by a linear encoder,
\begin{equation}\label{eqn5}
Z = X W_{\mathrm{enc}}^\top, \qquad W_{\mathrm{enc}} \in \mathbb{R}^{d \times n},
\end{equation}
producing embeddings \(Z \in \mathbb{R}^{m \times d}\). Next, we decode the embeddings with a bilinear low-rank decoder parametrised by factor matrices  \(U\in\mathbb{R}^{n\times r}\) and \(V\in\mathbb{R}^{d\times r}\):
\begin{equation}\label{eqn6}
W_{\mathrm{dec}} = U V^\top, \qquad
\hat{X} = Z W_{\mathrm{dec}}^\top = Z (U V^\top)^\top.
\end{equation}

The model parameters \(\theta=\{W_{\mathrm{enc}},U,V\}\) are optimized by minimizing the rank-aware objective
\begin{equation}\label{eqn7}
\mathcal{L}(\theta) = \frac{1}{|\Omega|}\|P_{\Omega}(\hat{X}-X)\|_F^2
+ \frac{\lambda_2}{2}\|Z\|_F^2
+ \lambda_1 \|Z\|_*,
\end{equation}
where the second controls embedding scale, and the third encourages low rank. Gradient descent updates on \(\theta\) are combined with periodic singular value thresholding (SVT) projections of $Z$ to maintain low-rank structure.

In this work, groups are treated as unordered sets of users \(g \subseteq \{1,\dots,m\}\) and are provided as part of the input data. Our focus is on modelling group preferences and interactions, given these
groupings, using low-rank representations and attention-based aggregation.

\noindent For group recommendation, let \(g\) be a group with member embeddings \(Z_g \in \mathbb{R}^{|g|\times d}\). We model non-linear group dynamics by passing \(Z_g\) through stacked Set Attention Blocks (SAB):
\begin{equation}
Z_g\ \rightarrow\ H^{(1)} \doteq \mathrm{SAB}_1(Z_g)
   \ \rightarrow\ H^{(2)} \doteq \mathrm{SAB}_2(H^{(1)})
   \ \rightarrow\ H^{(3)} \doteq \mathrm{SAB}_3(H^{(2)}).
\label{eq:attention}
\end{equation}
Pooling by Multihead Attention (PMA) replaces standard pooling operations, such as mean or max pooling, with a learnable, attention-based mechanism.
Given an input set of feature vectors $Z \in \mathbb{R}^{n \times d}$, PMA introduces $k$ learnable seed vectors $S \in \mathbb{R}^{k \times d}$ that act as queries attending over $Z$ \cite{Lee2019SetTransformer}. 
Formally, PMA is defined as
\begin{equation}
\mathrm{PMA}_k(Z) = \mathrm{MAB}(S, rrF(Z)),
\end{equation}
where $\mathrm{MAB}$ denotes a Multihead Attention Block. 
Each seed vector extracts a particular summary of the set, yielding $k$ pooled embeddings. 
When $k=1$, PMA produces a single unified representation suitable for group recommendation, while larger values of $k$ enable modelling tasks that require multiple correlated outputs. 

The Set-Transformer module serves as a group-level aggregation mechanism that operates on the latent user representations learned by the rank-constrained encoder-decoder. Individual user embeddings
$Z_i$ are learned and regularized independently through the low-rank RC-DMC objective and are used directly for individual-level prediction. For group recommendation, the Set-Transformer is applied only to the subset $Z_g$ corresponding to group members, producing a pooled group representation that is used for group-level prediction. In this way, attention augments the model by learning how to combine individual preferences into a coherent group representation, without modifying the underlying low-rank reconstruction procedure.

\paragraph{Set Transformer Components}
For completeness we recall the Set-Transformer building blocks \cite{FanChow2017} used in the group encoder. Let $\mathrm{Multihead}(Q,K,V)$ denote standard multi-head attention:
\begin{align}
\mathrm{Multihead}(Q,K,V) &= \mathrm{Concat}(O_1, \dots, O_h)\, W^O, \\
O_i &= \mathrm{softmax}\!\left(\frac{Q W_i^Q (K W_i^K)^\top}{\sqrt{d_k}}\right)\, V W_i^V,
\end{align}
where \(W_i^{Q,K,V}\) are per-head projections and \(W^O\) is the output projection. A basic block (MAB) with residual and feed-forward reads
\[
H = \mathrm{LayerNorm}\!\big(X + \mathrm{Multihead}(X,Y,Y)\big),\qquad
\mathrm{MAB}(X,Y) = \mathrm{LayerNorm}\!\big(H + \mathrm{rFF}(H)\big).
\]
From these, we define
\[
\mathrm{SAB}(X) = \mathrm{MAB}(X,X), \qquad
\mathrm{PMA}_k(X) = \mathrm{MAB}(S, X),
\]
where \(S\in\mathbb{R}^{k\times d}\) are learnable seed vectors. The Set-Transformer thus provides a permutation-invariant, learned aggregation that captures pairwise and higher-order interactions among group members.

\noindent We provide three related algorithmic components used in this work and clarify their relationships:

Group Soft-Impute (Alg.~\ref{algo1}) is an augmentation of the Soft-Impute algorithm of \cite{Mazumder2010}. Our Group Soft-Impute (previously presented on arXiv \cite{Ibrahim2025}) extends the atomic Soft-Impute by augmenting the data matrix with group-average rating rows (see Eq. ~\ref{eq:group_avg}) and associated per-item group weights (see Eq.~\ref{eq:group_weight})  used to form the augmented matrix. This extension modifies the input matrix via group-based augmentation, while retaining the standard Soft-Impute update structure.

RC-DMC: Training Loop (Alg.~\ref{algo2}) is our rank-constrained deep matrix completion training procedure. It is inspired by Deep Learning based Matrix Completion \cite{FanChow2017} and differs from a plain autoencoder in that (i) we parameterize the decoder as a low-rank factorization \(W_{\mathrm{dec}} = U V^\top\) and (ii) we interleave gradient updates on the model parameters \(\theta=\{W_{\mathrm{enc}},U,V\}\) with periodic SVT projections on the latent codes \(Z\) to explicitly enforce low-rank structure (see Eq.~\eqref{eqn7}). 

Group RC-DMC (Alg.~\ref{algo3}) combines the two components above with a Set-Transformer group encoder. Concretely, Group RC-DMC uses Group Soft-Impute to warm-start \(Z\) (Alg.~\ref{algo1}), trains the rank-constrained encoder/decoder via the RC-DMC loop (Alg.~\ref{algo2}), and then applies a Set-Transformer to per-group member embeddings \(Z_g\) to obtain unified group representations. The Set-Transformer addition is an architectural extension: it consumes individual embeddings and outputs pooled group representations (via learnable seeds) that are used only for group-level predictions.

\begin{algorithm}[t] 
\caption{WarmStart Initialization using Group Soft-Impute algorithm}
\label{algo1}
\begin{algorithmic}[1]
\Require Ratings matrix $X \in \mathbb{R}^{n\times m}$, observed entries $\Omega$, groups $\mathcal{G}$, thresholds $\tau_{\max}=\sigma_{\max}(P_\Omega(X))$, $\tau_{\min}=1$, convergence tolerance $\epsilon>0$
\Ensure Initial encoder $W_{\mathrm{enc}}$
\State Initialize $Z^{old} \gets 0$
\State Compute group-average ratings $r_{G}$ (as defined in Eq.~\eqref{eq:group_avg})
 \State Compute item weights $w_{G}$
 (as defined in Eq.~\eqref{eq:group_weight})
  \State Form augmented matrix $X_{\mathrm{aug}}$ (as defined in  Eq.~\eqref{eq:xaug})

\For{$\tau \in \{\tau_{\max}, \dots, \tau_{\min}\}$ decreasing}
\State Compute $U, D, V^\top \gets \mathrm{SVD}\big(P_\Omega(X_{aug}) + P_{\Omega^c}(Z^{old})\big)$
\State Apply singular-value soft-thresholding to $D$ (as defined in Eq.~\eqref{eq:svt_def})
\State Update $Z^{new} \gets U D_\tau V^\top$
\If{$\dfrac{\|Z^{new}-Z^{old}\|_F}{\|Z^{old}\|_F} < \epsilon$}
        \State \textbf{break}
    \EndIf
    \State $Z^{old} \gets Z^{new}$
\EndFor
\State Compute encoder by least-squares: $W_{\mathrm{enc}} \gets \mathrm{lstsq}(X_{aug}, Z^{new})^\top$
\State \Return $W_{\mathrm{enc}}$
\end{algorithmic}
\end{algorithm}

\begin{algorithm}[t]
\caption{RC-DMC: Training Loop (Rank-Constrained Matrix Completion)}
\label{algo2}
\begin{algorithmic}[1]
\Require Ratings matrix $X$, mask $\Omega$, groups $\mathcal{G}$, latent dimensions $d,r$, learning rate $\eta$, regularization weights $\lambda_1,\lambda_2$, epochs $T$.
\Ensure Trained parameters $\theta = \{W_{\mathrm{enc}},U,V\}$ and reconstruction $\widehat{X}$
\State Initialize $W_{\mathrm{enc}} \gets \text{WarmStart}(X,\Omega,\mathcal{G},r,d)$
\State Initialize $U \in \mathbb{R}^{n \times r}, \; V \in \mathbb{R}^{d \times r}$ randomly
\For{epoch $e = 1,\dots,T$}
  \State Encode latent codes: $Z \gets \text{encode}(X, W_{\mathrm{enc}})$
  \State Decode reconstruction: $\widehat{X} \gets \text{decode}(Z, U, V)$
  \State Compute loss:
  \[
    \mathcal{L} = \frac{\| P_\Omega(X-\widehat{X}) \|_F^2}{|\Omega|} + \lambda_2 \|Z\|_F^2 + \lambda_1 \|Z\|_*
  \]
  \State Update parameters $\theta \gets \theta - \eta \, \nabla_\theta \mathcal{L}$ 
  \State Apply SVT: $Z' \gets \mathrm{SVT}(Z)$
  \State Recompute encoder: $W_{\mathrm{enc}} \gets \text{lstsq}(X,Z')^\top$
\EndFor
\State \Return $\theta, \widehat{X}$
\end{algorithmic}
\end{algorithm}

\begin{algorithm}[t]
\caption{Group RC-DMC}
\label{algo3}
\label{alg:group-rcdmc}
\begin{algorithmic}[1]
\Require Trained RC-DMC parameters $\theta=\{W_{\mathrm{enc}},U,V\}$, groups $\mathcal{G}$
\Ensure Group predictions $\{\hat{x}_g : g \in \mathcal{G}\}$
\For{each group $g \in \mathcal{G}$}
    \State Encode members: $Z_g \gets \{x_u W_{\mathrm{enc}}^\top : u \in g\}$
    \State Apply stacked SAB layers: $H^{(3)} \gets \mathrm{SAB}(\mathrm{SAB}(\mathrm{SAB}(Z_g)))$
    \State Aggregate via PMA: $z_g \gets \mathrm{PMA}(H^{(3)})$
    \State Decode ratings: $\hat{x}_g \gets z_g (UV^\top)^\top + b_g$
\EndFor
\State \Return $\{\hat{x}_g\}$
\end{algorithmic}
\end{algorithm}
\FloatBarrier

\subsection{Computational Complexity and Convergence}

Group RC-DMC involves three main computational building blocks:(i) encoding the observed
user–item ratings into latent codes, (ii) reconstructing predictions via the low-rank decoder and
occasional singular-value thresholding (SVT), and (iii) group processing via the Set-Transformer.
Below we summarise the dominant costs and the convergence behaviour of the complete procedure.

The proposed RC-DMC model achieves computational efficiency by operating directly on the observed entries of the user–item rating matrix, computing each user’s latent representation solely from their observed ratings. 
For the purpose of complexity analysis, we use the encoder and decoder definitions introduced in Section~\ref{sec:methodology} (see Eq.~\eqref{eqn5} and Eq.~\eqref{eqn6}).
Forming all user embeddings by applying the encoder to the observed entries requires \(\mathcal{O}(|\Omega|\,d)\) operations, where \(|\Omega|\) is the number of observed user-item pairs and \(d\) the embedding dimension.
Reconstructing a single user–item entry via the low-rank decoder parameterized as \(W_{\mathrm{dec}}=UV^\top\) costs \(\mathcal{O}(d)\), hence evaluating the reconstruction on all observed entries incurs \(\mathcal{O}(|\Omega|\,d)\) work.

In total, when the decoder matrix is \emph{constructed in full}, the dominant 
computational cost per epoch is
\[
\mathcal{O}\big(|\Omega|\,d + nrd\big),
\]
where the $\mathcal{O}(nrd)$ term is usually incurred only once per epoch.

Periodic SVT projections are applied to the encoder output matrix $Z\in\mathbb{R}^{m\times d}$ to enforce and maintain a low-rank latent structure,  a truncated SVD to rank $r$ requires $\mathcal{O}(m d r)$ operations.
When SVT is performed every $K_{\mathrm{SVT}}$ iterations, its amortized contribution per iteration is $\mathcal{O}\!\left(\frac{m d r}{K_{\mathrm{SVT}}}\right)$. 
Thus, including the encoder and decoder computations as well as the amortized cost of periodic SVT, the practical per-iteration complexity becomes
$$
\mathcal{O}\!\Big(|\Omega|\,d \;+\; n r d \;+\; \frac{m d r}{K_{\mathrm{SVT}}}\Big).
$$

For group recommendation, RC-DMC augments per-user latent factors with a Set-Transformer encoder applied to each group \(g\in\mathcal{G}\). Self-attention among group members forms an \(|g|\times|g|\) attention matrix: for a single head the dominant cost is \(\mathcal{O}(|g|^2 d)\), and with \(h\) heads and \(L\) stacked Set-Attention Blocks (SAB) the per-group cost is
$$
    \mathcal{O}\!\big(L\,h\,|g|^2 d\big).
$$
Pooling-by-Multihead-Attention (PMA) with \(k\) seed vectors adds \(\mathcal{O}(k\,|g|\,d)\) (plus the small seed-wise \(\mathcal{O}(k^2 d)\) term). Combining encoder, decoder, amortized SVT and the aggregate group attention cost, the practical per-epoch complexity of Group RC-DMC is
\[
\mathcal{O}\!\Big(|\Omega|\,d \;+\; n r d \;+\; \frac{m d r}{K_{\mathrm{SVT}}}
\;+\; \sum_{g\in\mathcal{G}} L h |g|^2 d\Big),
\]
where $|\Omega|$ is the number of observed entries, $d$ the encoder dimension, $r$ the decoder rank, $K_{\mathrm{SVT}}$ the SVT scheduling interval, and the $\mathcal{O}(n r d)$ term corresponds to forming the explicit decoder matrix $W_{\mathrm{dec}}=U V^\top$ (this term is typically paid once per epoch and is therefore effectively amortized). 
Multiplying by the number of training epochs $T$ gives the total training cost; add the one-time warm-start cost $\mathcal{O}(I_{\mathrm{SI}}\,m'\,n\,r)$ (where $m'$ is the row count of the augmented matrix and $I_{\mathrm{SI}}$ is the number of Soft-Impute iterations).

\subsubsection{Convergence}
We emphasize two complementary convergence statements that apply to different
parts of our pipeline.
The Soft-Impute warm-start applied to the augmented matrix $X_{\mathrm{aug}}$
solves the convex nuclear-norm problem in Eq.~\eqref{eqn3}. As shown in
\cite{Mazumder2010}, the iterative SVT procedure
used by Soft-Impute produces a monotonically decreasing objective and converges
to the global minimizer of the nuclear-norm regularized problem; in particular,
the relative iterate gap $\|Z^{(k+1)}-Z^{(k)}\|_F / \|Z^{(k)}\|_F$ tends to zero.

The joint training of RC-DMC (encoder parameters $W_{\mathrm{enc}}$ and bilinear
decoder factors $U,V$) minimizes the composite objective
\[
\mathcal{L}(\theta) \;=\; g(\theta) + \lambda\|Z(\theta)\|_*,
\qquad \theta=\{W_{\mathrm{enc}},U,V\},
\]
where $g(\theta)$ denotes the smooth mean-squared error plus any Frobenius
penalties, and $Z(\theta)$ is the encoder output matrix.
This objective is non-convex due to the bilinear decoder and encoder parametrisation.
In practice we perform gradient  updates on the smooth term $g$ and apply SVT as the proximal operator of the nuclear norm penalty on $Z$.
Under standard regularity conditions, namely bounded iterates, a continuous Lipschitz gradient of $g$, and suitably chosen step sizes, the proximal gradient theory for non-convex composite problems
guarantees that every limit point of the generated sequence is a critical point of $\mathcal{L}$ \cite{Li2015}.
Global optimality cannot be guaranteed because of non-convexity.
Empirically we observe that the training loss decreases smoothly and plateaus, and that initializing from Soft-Impute substantially reduces the number of gradient iterations required and improves robustness to poor local minima.
\subsection{Baseline}

We evaluate our proposed method by comparing it against two established techniques: the weighted before factorisation method and the after factorisation method \citep{Ortega2016}. Unlike the soft-impute algorithm, these group recommendation methods are derived MF approaches.  

The MF-based group recommendation framework is extended in two ways: 
\begin{enumerate}
    \item Ratings for a selected group of users are first aggregated using an aggregation function $\mathbf{h}$ to construct a pseudo-group user $u_g$, before applying factorization to the rating matrix. 
    \item Ratings are computed after aggregating predictions derived from the latent factor matrices.  
\end{enumerate}
These two techniques serve as strong baselines for evaluating group recommendation systems within collaborative filtering frameworks. By effectively capturing and integrating user preferences, they provide a robust benchmark for assessing the performance of the proposed approach.  

In the following sections, we present a detailed description of these methods.
\subsubsection*{After Factorisation Method}
The after factorisation (AF) approach proposed in the literature \citep{Ortega2016} recommends items to a group of users by leveraging the latent factors learned through MF. Initially, MF model is trained to obtain latent factor representations for both users and items, effectively capturing their underlying preferences and characteristics.  

To model group preferences, the individual latent factors of the group members are aggregated into a group profile $\mathbf{u}_G$ and item profile $\mathbf{v}_G$, respectively, using a predefined aggregation function $\mathbf{h}$.

Commonly used aggregation functions include the average, weighted average, minimum aggregation and maximum aggregation.

\subsubsection*{Weighted Before Factorization Method }

The weighted before factorization (WBF) approach \cite{Ortega2016} focuses on computing a weighted aggregation of the individual preferences in the user rating matrix before applying MF. By emphasizing the importance of certain entries prior to factorization, this method constructs a collective preference profile for the group while ensuring a degree of fairness among its members.

For a group \( G \) with a set of items, weights \( w_{G,j} \) are defined as in equation \eqref{eq:group_weight}
The group user  profile \( \mathbf{u}_G \) is computed by solving the objective function described in equation~\eqref{eq:mmmf}.
\begin{equation}\label{eq:mmmf}
    \begin{aligned}
        \min_{\mathcal{U}, \mathcal{V}} &\quad || P_\Omega (\mathbf{X}) - P_\Omega (\mathcal{U}\mathcal{V}^T) ||^2_F & + \lambda (||\mathcal{U}||^2_F + ||\mathcal{V}||^2_F)
    \end{aligned}
\end{equation}

\section{Experiments}\label{sec:Experiments}

\subsection{Datasets}\label{sec:Datasets}

We evaluate Group RC-DMC on two widely used real-world recommender datasets: MovieLens \cite{Harper2015} and Goodbooks \cite{Goodbooks2017}.

\begin{table}[!h]
\caption{The Movielens (left) and the GoodBooks (right) datasets. }\label{tab:datasets}%
\begin{tabularx}{.7\textwidth}{@{}l|l@{}l@{}l|l}
\toprule
Movielens & Details  &{\qquad\qquad\qquad} & Goodbooks & Details \\
\midrule
\#number of users   & 943  & {~} & \#number of users    & 53K \\
\hline
\#number of movies  & 1682 & {~} & \#number of books    & 10K \\
\hline
\#number of ratings & 100K & {~} & \#number of ratings  & 6 million \\
\botrule
\end{tabularx} %
\end{table}

The MovieLens dataset (Table~\ref{tab:datasets} left) contains 943 users and 1,682 items with 100{,}000 ratings on a 1-5 scale and a sparsity of approximately 86\%. For the experiments reported in this work we extract a subset of size \(943\times 500\) (users \(\times\) items) for computational purposes. The observed entries are randomly partitioned into training and test sets, comprising roughly 80\% and 20\% of the observed ratings, respectively.
The Goodbooks dataset (Table~\ref{tab:datasets} right) is a large-scale book-rating dataset containing millions of ratings across tens of thousands of users and items, with an overall sparsity of approximately \(86\%\). For experiments we extract a representative subset of size \(1000\times 500\). Ratings are on a 1–5 scale, and the observed entries are split 80\%/20\% into training and test sets.

All experiments report results on held-out observed ratings (test set) and use the same masking/splitting procedure for both datasets to ensure comparable evaluation. 

\subsection{Experimental Results}

We evaluate Group RC-DMC on two real-world datasets (MovieLens and Goodbooks) using the train/test splits described in subsection \ref{sec:Datasets}. Our experimental evaluation focuses on two complementary aspects: (i) group recommendation quality and (ii) rating reconstruction accuracy.

\noindent Group recommendation quality is evaluated using precision, recall, and F$_1$ measure; all these computed from aggregated group ratings. Specifically, for each group $G$ and item $i$, the aggregated true group rating is defined as
\[
r(G,i) = \frac{1}{|G|}\sum_{u \in G} x_{u,i},
\]
while the corresponding aggregated group prediction is computed similarly but using the predictions $\hat{x}_i$.
These metrics are computed per group, following prior work \cite{Felfernig2018book}. In what follows we define \emph{precision} and \emph{recall} \citep{Ting2016}: 

\noindent \emph{Precision} is the proportion of correctly retrieved \emph{relevant} items, calculated as the ratio of relevant recommended items to the total number of recommended items. 

\noindent \emph{Recall} is the ratio of the number of relevant recommended items to the total number of relevant items in the dataset.

\noindent \emph{F1 score} \citep{Hand2021} is the harmonic mean of precision and recall.
The definitions are:
\begin{equation}\label{eq:prec_recall}
  \mathrm{precision = \frac{TP}{TP+FP}},
  \quad 
  \mathrm{recall = \frac{TP}{TP+FN}},
  \quad 
  \text{F1} = 2 \cdot \frac{\mathrm{precision} \cdot \mathrm{recall}}{\mathrm{precision} + \mathrm{recall}}
\end{equation}
where TP, FP, and FN denote true positives, false positives, and false negatives, respectively.
Precision, recall, and F1 are computed for each group by thresholding the predicted and true aggregated group ratings over all items (threshold $t=3.5$).

\noindent In addition to group-level recommendation quality, we evaluate the reconstruction accuracy of the predicted ratings using the root mean squared error (RMSE) on both training and test sets:

\begin{equation*}
   \mathrm{RMSE} \;=\;
   \sqrt{
   \frac{P_{\Omega}(X - \hat{X})^2}{|\Omega|}}.
\end{equation*}

\noindent Following the evaluation framework for group recommender systems
introduced by ~\cite{Trattner2024}, we evaluate group-level
prediction accuracy by computing an error metric separately for each group and
then averaging across groups. While Trattner et al. employ mean absolute error
(MAE), we instead use RMSE to remain consistent with
the atomic reconstruction objective used throughout this paper.

Let $\Omega_G$ denote the set of items for which at least one member of group
$G$ has an observed rating in the test set. The group RMSE for group $G$ is
defined as
\begin{equation*}
\mathrm{RMSE}(G) =
\sqrt{
\frac{1}{|\Omega_G|}
\sum_{i \in \Omega_G}
\bigl(r(G,i) - \hat{r}(G,i)\bigr)^2 }.
\end{equation*}

The overall group RMSE is obtained by averaging $\mathrm{RMSE}(G)$ across all
groups.

\begin{figure}[t]
  \centering
  \begin{tabular}{ccc}
    \tikz\node[rectangle,outer sep=0, inner sep=0,rotate=90]{\qquad \quad MovieLens results};
    &
    \includegraphics[width=.45\textwidth]{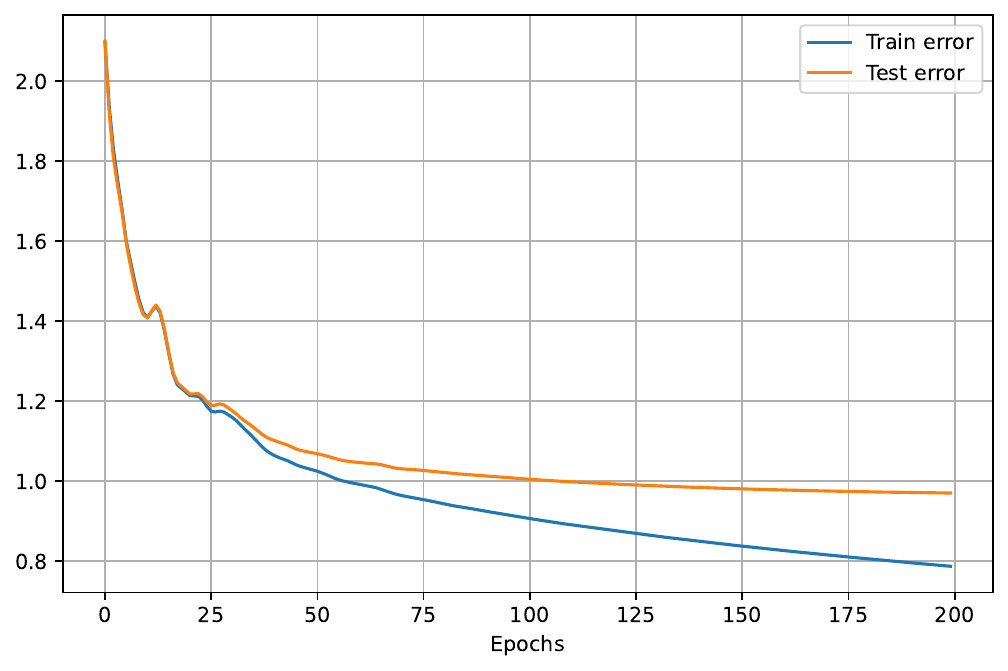} 
    &
    \includegraphics[width=.45\textwidth]{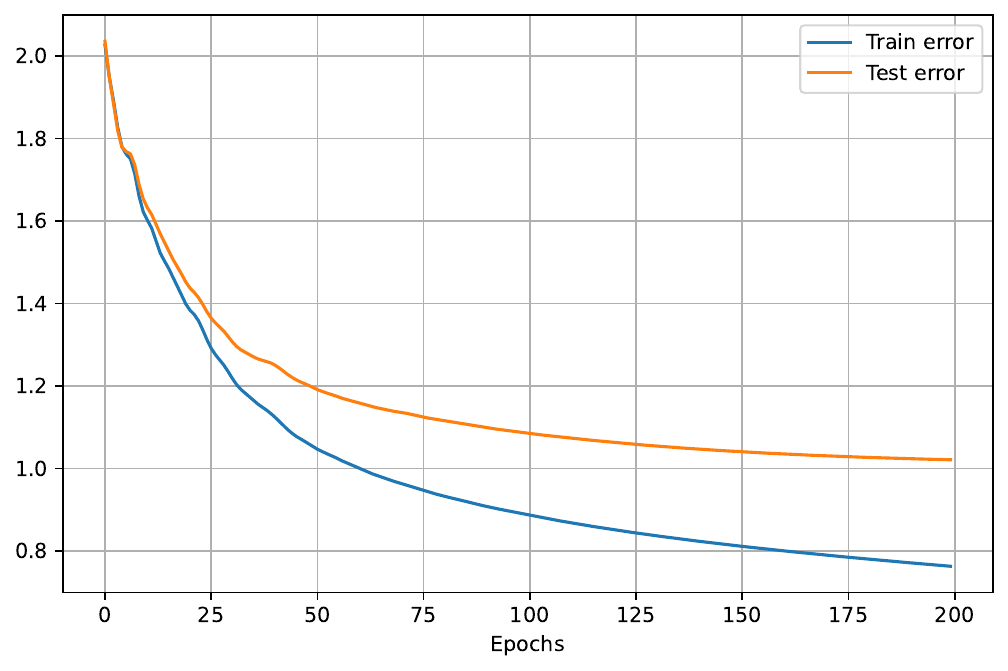}
    \\
    {} & Dense Encoder & Sparse Encoder
    \\
    \tikz\node[rectangle,outer sep=0, inner sep=0,rotate=90]{GoodBooks results};
    &
    \includegraphics[width=.45\textwidth]{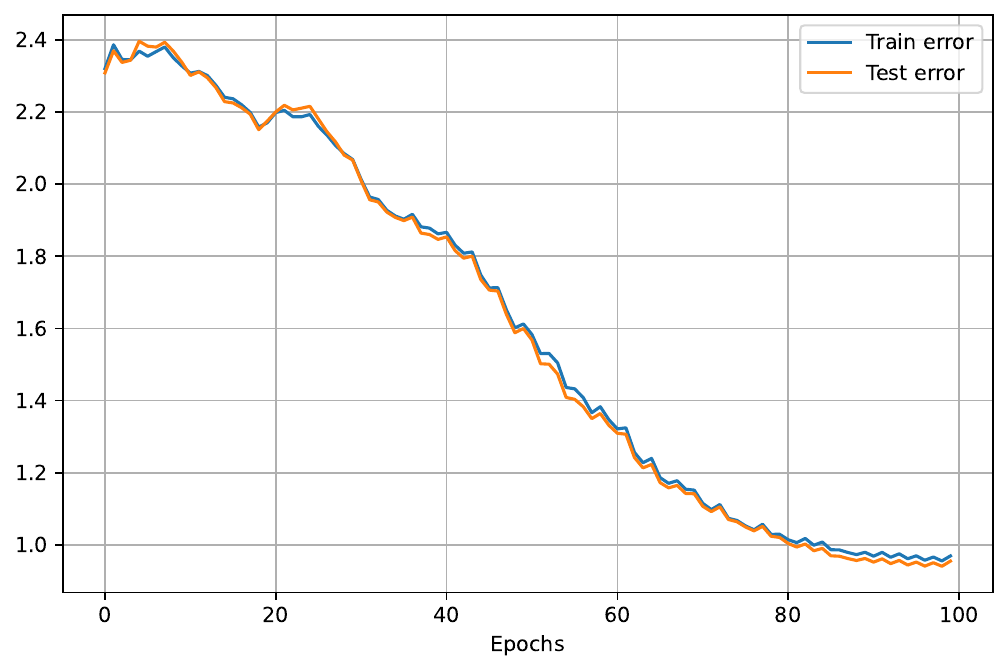}
    &
    \includegraphics[width=.45\textwidth]{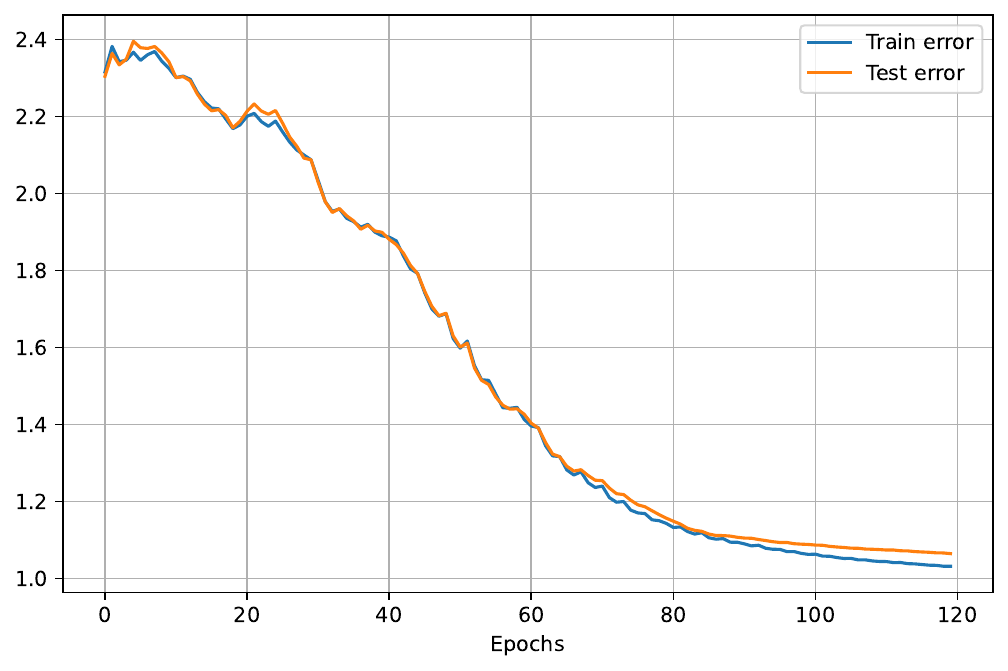}
    \\
    {} & Dense Encoder & Sparse Encoder
\end{tabular}
\caption{
Training and test RMSE comparison for dense and sparse RC-DMC encoders on the MovieLens and GoodBooks datasets. 
The sparse encoder converges faster during early epochs, while the dense encoder achieves lower final RMSE, indicating more stable and accurate reconstruction performance across datasets.%
}
\label{fig:rmse_all}
\end{figure}

\begin{figure}[t]
  \centering
  \includegraphics[width=1.0\textwidth]{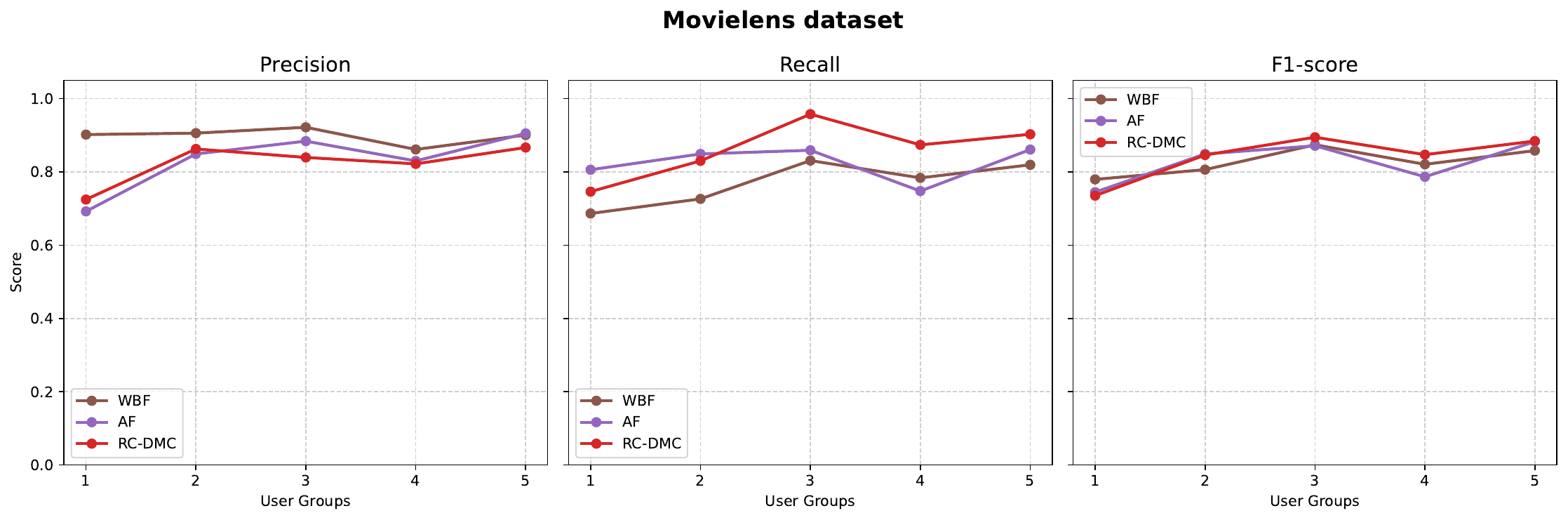}
  \caption{MovieLens: group-level Precision / Recall / F1 comparison between RC-DMC and two baselines (WBF, AF)}
  \label{fig:movielens-prf}
\end{figure}

\begin{figure}[t]
  \centering
  \includegraphics[width=1.0\textwidth]{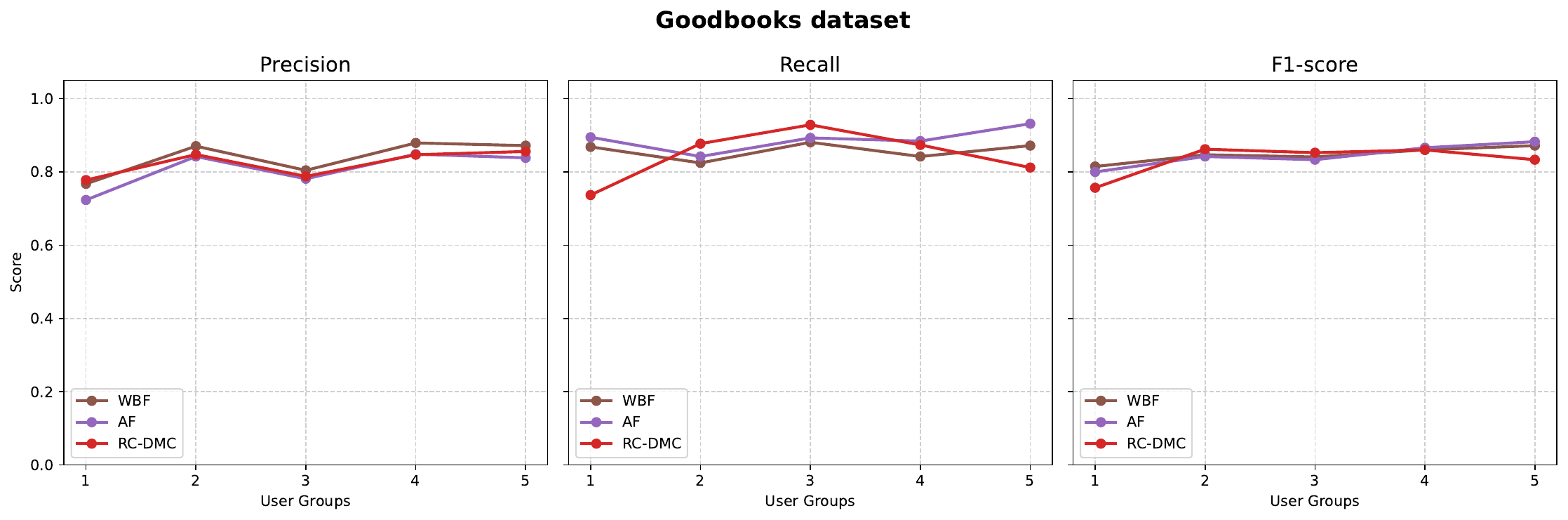}
  \caption{Goodbooks: group-level Precision / Recall / F1 comparison between RC-DMC and baselines (WBF, AF)}
  \label{fig:goodbooks-prf}
\end{figure}
To evaluate the proposed method (Group RC-DMC), we compared it with baselines (WBF and AF approach).
The evaluation was based on precision, recall, and F1 score computed for all recommended items. Figure~\ref{fig:movielens-prf} and Figure~\ref{fig:goodbooks-prf} illustrate the results across user groups of varying sizes (5, 10, 15, 20, and 25).

For the Movielens dataset, Group RC-DMC achieves the highest recall among the three methods, averaging 0.86 across groups, which indicates its strength in capturing relevant items. However, this comes at the cost of lower precision (0.82), meaning it tends to recommend more false positives compared to WBF and AF. WBF, by contrast, achieves the highest precision ($\sim$ 0.90) but suffers from lower recall (0.75), suggesting it is more conservative in its recommendations. AF provides a balanced performance, with precision and recall both averaging around 0.83, resulting in an F1 score of 0.83. Group-wise, RC-DMC excels in recall for groups 3, 4 and 5, while WBF dominates in precision for groups 1, 2, 3 and 4. AF maintains stable performance across all groups, without extreme highs or lows, as shown in Figure~\ref{fig:movielens-prf}.

For the Goodbooks dataset, the results indicate that RC-DMC achieves precision values comparable to those of WBF and AF across all groups, with WBF performing slightly better on average. This indicates that while RC‑DMC remains competitive, WBF shows a slight advantage in reducing false positives, resulting in marginally more relevant recommendations. In terms of recall, RC-DMC achieves competitive performance, particularly for groups 2 and 3, where it reaches recall values above 0.92. However, AF also demonstrates strong recall across most groups, with values consistently above 0.84, while WBF maintains a balanced trade-off between precision and recall. The F1 scores reflect this balance: RC-DMC achieves an average F1 of 0.83, while WBF and AF slightly outperform it with averages of 0.85. Overall, RC‑DMC demonstrates strong performance across groups, particularly excelling in recall, which highlights its ability to capture a larger proportion of relevant items while maintaining competitive precision and F1 scores, as illustrated in Figure~\ref{fig:goodbooks-prf}.

\noindent Therefore, the method used in this paper significantly enhances the quality of recommendations in terms of precision and recall for small, mid size and large user groups.

\noindent The RMSE training plot from Fig.~\ref{fig:rmse_all} shows that the warm-start plus rank-constrained training produce rapid early improvements followed by steady convergence consistent with the intended effect of the Soft-Impute initialization. 

 The sparse encoder computes each user's latent vector using only their observed ratings, giving complexity \(\mathcal{O}(|\Omega|d)\).

Group metrics demonstrate that RC-DMC captures group preferences effectively: in many groups RC-DMC attains F1-scores comparable to or exceeding the baselines, while in some groups baselines achieve slightly higher precision at the cost of recall. These differences are expected: WBF strategies emphasize conservative recommendations (high precision), whereas RC-DMC’s learned aggregation balances precision and recall by modelling within-group interactions via the Set-Transformer.

Table~\ref{tab:main-results} complements these results by reporting the training efficiency and group RMSE. AF and WBF were retrained per group, resulting in higher total training times, whereas RC-DMC is trained once and reused for all groups, making it substantially faster than WBF. AF remains faster than both due to its simple matrix factorization approach, RC-DMC achieves the lowest group RMSE, demonstrating superior reconstruction accuracy. These results illustrate that RC-DMC achieves a favorable trade-off between accuracy and efficiency, providing competitive reconstruction quality and group-level recommendation performance at reduced computational cost.

\begin{table}[t]
\centering
\caption{Group RMSE and total training time. AF and WBF are evaluated under a retrain-per-group protocol,
and their reported training time is the sum across all groups.}
\label{tab:main-results}
\begin{tabular}{lcc|cc}
\toprule
 & \multicolumn{2}{c}{\textbf{MovieLens}} &
 \multicolumn{2}{c}{\textbf{Goodbooks}} \\
\cmidrule(lr){2-3} \cmidrule(lr){4-5}
\textbf{Method} &
Group RMSE  & Train time (s) &
Group RMSE  & Train time (s) \\
\midrule
RC-DMC  & 0.9706 & 41.03 & 0.8936 & 44.61 \\
AF  & 1.0514 & 5.79  & 0.9981 & 5.96  \\
WBF & 1.0018 & 64.38 & 0.9458 & 67.54 \\
\bottomrule
\end{tabular}
\end{table}

In summary, on both MovieLens and GoodBooks the Group RC-DMC model attains strong reconstruction accuracy and competitive group-level recommendation quality. The warm-start (Soft-Impute) initialization shortens the transient phase of training and improves robustness.

\section{Conclusion}
We introduced a novel group recommendation method, Group RC-DMC, which unifies low-rank matrix completion with attention-based group modelling. The framework combines an explicit low-rank prior (Soft-Impute warm-start and nuclear-norm regularisation), efficient linear encoder–decoder mappings, and a Set transformer for non-linear group aggregation. 
Empirically, Group RC-DMC attains performance comparable to strong MF baselines (AF and WBF) and yields small but consistent improvements in F$1$ under our experimental settings.

Crucially, Group RC-DMC provides several complementary benefits that justify its use beyond Group MF baselines. 
First, the nuclear-norm regulariser and periodic SVT enforce a rank-controlled latent structure, yielding compact representations that are easier to interpret and compress than the unconstrained latent factors produced by standard MF models. Second, Set transformer based aggregation captures pairwise and higher-order interactions among group members that simple aggregation rules cannot model, improving recommendation quality for groups with diverse interests.

Finally, from a computational perspective, Group RC-DMC’s dominant per-epoch operations scale with the number of observed entries and with a controllable latent dimension, while the cost of truncated SVD used for occasional SVT is amortised across iterations. In contrast, classical MF-based group methods such as AF and WBF scale with the MF factor dimension and with the size of the group item set \cite{Ortega2016}.
The results indicate that the proposed approach effectively recovers the underlying low-rank structure of the rating matrix and yields high-quality group recommendations across small, medium and large groups. 
For future work we will augment Group RC-DMC with side information (demographics, item metadata and textual embeddings) to better address cold-start scenarios. While the high-dimensional, partially observed matrices used here mitigate cold-start in many practical settings, explicit cold-start mechanisms remain important; we plan to explore integrating such strategies directly into the Group RC-DMC pipeline.

\backmatter

\section*{Declarations}

\begin{flushleft}

\textbf{Author Contributions} Not applicable.\\[0.2em]
\textbf{Funding} Not applicable.\\[0.2em]
\textbf{Data Availability}
The datasets used in this study are publicly available. The \textit{MovieLens} dataset can be accessed at \url{https://doi.org/10.1145/2827872}, and the \textit{Goodbooks-10k} dataset is available at \url{http://fastml.com/goodbooks-10k} respectively.\\[0.2em]
\textbf{Conflict of interest} On behalf of all authors, the corresponding author states that there is no conflict of interest.\\[0.2em]
\textbf{Research Involving Human and/or Animals} Not applicable.\\[0.2em]
\textbf{Informed Consent} Not applicable.
\end{flushleft}

\backmatter
\bibliography{sn-bibliography}

\begin{thebibliography}{10}
\providecommand{\doi}[1]{\url{https://doi.org/#1}}
\bibcommenthead

\bibitem[\protect\citeauthoryear{Cand{\`e}s and Tao}{2010}]{CandesTao2009}
Cand{\`e}s EJ, Tao T.
\newblock The Power of Convex Relaxation: Near-Optimal Matrix Completion.
\newblock IEEE Transactions on Information Theory. 2010;56(5):2053--2080.
\newblock \doi{10.1109/TIT.2010.2044061}.

\bibitem[\protect\citeauthoryear{Keshavan et~al.}{2010}]{Keshavan2010}
Keshavan RH, Montanari A, Oh S.
\newblock Matrix Completion From a Few Entries.
\newblock IEEE Transactions on Information Theory. 2010;56(6):2980--2998.
\newblock \doi{10.1109/TIT.2010.2046205}.

\bibitem[\protect\citeauthoryear{He et~al.}{2017}]{He2017}
He X, Liao L, Zhang H, Nie L, Hu X, Chua TS.
\newblock Neural Collaborative Filtering.
\newblock In: Proceedings of the 26th International Conference on World Wide Web. WWW '17. Republic and Canton of Geneva, CHE: International World Wide Web Conferences Steering Committee; 2017. p. 173–182.

\bibitem[\protect\citeauthoryear{Jameson and Smyth}{2007}]{jameson2007}
Jameson A, Smyth B.
\newblock Recommendation to Groups.
\newblock In: Brusilovsky P, Kobsa A, Nejdl W, editors. The Adaptive Web: Methods and Strategies of Web Personalization. Berlin, Heidelberg: Springer; 2007. p. 596--627.

\bibitem[\protect\citeauthoryear{Li et~al.}{2024}]{Li2024survey}
Li Y, Liu K, Satapathy R, Wang S, Cambria E.
\newblock Recent Developments in Recommender Systems: A Survey [Review Article].
\newblock IEEE Computational Intelligence Magazine. 2024;19(2):78--95.
\newblock \doi{10.1109/MCI.2024.3363984}.

\bibitem[\protect\citeauthoryear{Ortega et~al.}{2016}]{Ortega2016}
Ortega F, Hernando A, Bobadilla J, Kang JH.
\newblock Recommending items to group of users using Matrix Factorization based Collaborative Filtering.
\newblock Information Sciences. 2016;345:313--324.
\newblock \doi{10.1016/j.ins.2016.01.083}.

\bibitem[\protect\citeauthoryear{Fan and Chow}{2017}]{FanChow2017}
Fan J, Chow TWS.
\newblock Deep learning based matrix completion.
\newblock Neurocomputing. 2017;266:540--549.
\newblock \doi{10.1016/j.neucom.2017.05.074}.

\bibitem[\protect\citeauthoryear{Nguyen et~al.}{2018}]{Nguyen2018}
Nguyen DM, Tsiligianni E, Calderbank R, Deligiannis N.
\newblock Regularizing Autoencoder-Based Matrix Completion Models via Manifold Learning.
\newblock In: 2018 26th European Signal Processing Conference (EUSIPCO); 2018. p. 1880--1884.

\bibitem[\protect\citeauthoryear{Netflix}{2006}]{netflixprize}
Netflix.: Netflix Prize Dataset.
\newblock Over 100 million ratings from 480,189 users on 17,770 movies.
\newblock Available at \url{http://www.netflixprize.com/}.

\bibitem[\protect\citeauthoryear{Amazon}{2018}]{amazonreviews}
Amazon.: Amazon Product Reviews Dataset.
\newblock Amazon product reviews and metadata.
\newblock \url{https://nijianmo.github.io/amazon/index.html}.

\bibitem[\protect\citeauthoryear{Yelp}{2015}]{yelpopen}
Yelp.: Yelp Open Dataset.
\newblock User reviews, business metadata, and check-ins.
\newblock \url{https://www.yelp.com/dataset}.

\bibitem[\protect\citeauthoryear{Cai et~al.}{2010}]{Cai2010}
Cai JF, Cand\`{e}s EJ, Shen Z.
\newblock A Singular Value Thresholding Algorithm for Matrix Completion.
\newblock SIAM Journal on Optimization. 2010;20(4):1956--1982.
\newblock \doi{10.1137/080738970}.

\bibitem[\protect\citeauthoryear{Mazumder et~al.}{2010}]{Mazumder2010}
Mazumder R, Hastie T, Tibshirani R.
\newblock Spectral Regularization Algorithms for Learning Large Incomplete Matrices.
\newblock Journal of Machine Learning Research. 2010;11:2287--2322.

\bibitem[\protect\citeauthoryear{Cand\`{e}s and Plan}{2009}]{CandesPlan2009}
Cand\`{e}s EJ, Plan Y.
\newblock Accurate low-rank matrix recovery from a small number of linear measurements.
\newblock In: 2009 47th Annual Allerton Conference on Communication, Control, and Computing (Allerton); 2009. p. 1223--1230.

\bibitem[\protect\citeauthoryear{Koren et~al.}{2009}]{KorenBellVolinsky2009}
Koren Y, Bell R, Volinsky C.
\newblock Matrix factorization techniques for recommender systems.
\newblock Computer. 2009;42(8):30--37.

\bibitem[\protect\citeauthoryear{Cand{\`e}s and Recht}{2009}]{Emmanuel2009}
Cand{\`e}s EJ, Recht B.
\newblock Exact Matrix Completion via Convex Optimization.
\newblock Foundations of Computational Mathematics. 2009;9(6):717--772.
\newblock \doi{10.1007/s10208-009-9045-5}.

\bibitem[\protect\citeauthoryear{Hu et~al.}{2008}]{Hu2008}
Hu Y, Koren Y, Volinsky C.
\newblock Collaborative Filtering for Implicit Feedback Datasets.
\newblock In: 2008 Eighth IEEE International Conference on Data Mining; 2008. p. 263--272.

\bibitem[\protect\citeauthoryear{Bengio et~al.}{2006}]{Bengio2006}
Bengio Y, Lamblin P, Popovici D, Larochelle H.
\newblock Greedy layer-wise training of deep networks.
\newblock In: Proceedings of the 20th International Conference on Neural Information Processing Systems. NIPS'06. Cambridge, MA, USA: MIT Press; 2006. p. 153–160.

\bibitem[\protect\citeauthoryear{Lee et~al.}{2019}]{Lee2019SetTransformer}
Lee J, Lee Y, Kim J, Kosiorek A, Choi S, Teh YW.
\newblock Set Transformer: A Framework for Attention-based Permutation-Invariant Neural Networks.
\newblock In: Proceedings of the 36th International Conference on Machine Learning. vol.~97 of Proceedings of Machine Learning Research. PMLR; 2019. p. 3744--3753.
\newblock Available from: \url{https://proceedings.mlr.press/v97/lee19d.html}.

\bibitem[\protect\citeauthoryear{Ibrahim et~al.}{2025}]{Ibrahim2025}
Ibrahim MS, Saidu IC, Csato L.: Enhancing Group Recommendation using Soft Impute Singular Value Decomposition.
\newblock Available from: \url{https://arxiv.org/abs/2511.11172}.

\bibitem[\protect\citeauthoryear{Li and Lin}{2015}]{Li2015}
Li H, Lin Z.
\newblock Accelerated proximal gradient methods for nonconvex programming.
\newblock In: Proceedings of the 29th International Conference on Neural Information Processing Systems - Volume 1. NIPS'15. Cambridge, MA, USA: MIT Press; 2015. p. 379–387.

\bibitem[\protect\citeauthoryear{Harper and Konstan}{2015}]{Harper2015}
Harper FM, Konstan JA.
\newblock The MovieLens Datasets: History and Context.
\newblock ACM Trans Interact Intell Syst. 2015 Dec;5(4).
\newblock \doi{10.1145/2827872}.

\bibitem[\protect\citeauthoryear{Zajac}{2017}]{Goodbooks2017}
Zajac Z.: Goodbooks-10k: a new dataset for book recommendations.
\newblock FastML.
\newblock \url{http://fastml.com/goodbooks-10k}.

\bibitem[\protect\citeauthoryear{Felfernig et~al.}{2018}]{Felfernig2018book}
Felfernig A, Boratto L, Stettinger M, Tkal{\v{c}}i{\v{c}} M.
\newblock Evaluating Group Recommender Systems.
\newblock In: Group Recommender Systems: an Introduction. Cham: Springer International Publishing; 2018. p. 59--71.

\bibitem[\protect\citeauthoryear{Ting}{2016}]{Ting2016}
Ting KM.
\newblock Precision and Recall.
\newblock In: Sammut C, Webb GI, editors. Encyclopedia of Machine Learning and Data Mining. Boston, MA: Springer; 2016. p. 1--1.

\bibitem[\protect\citeauthoryear{Hand et~al.}{2021}]{Hand2021}
Hand DJ, Christen P, Kirielle N.
\newblock {F}*: an interpretable transformation of the {F}-measure.
\newblock Machine Learning. 2021;110(3):451--456.
\newblock \doi{10.1007/s10994-021-05964-1}.

\bibitem[\protect\citeauthoryear{Trattner et~al.}{2024}]{Trattner2024}
Trattner C, Said A, Boratto L, Felfernig A.
\newblock Evaluating Group Recommender Systems.
\newblock In: Felfernig A, Boratto L, Stettinger M, Tkalčič M, editors. Group Recommender Systems. Signals and Communication Technology. Cham: Springer; 2024. .

\end{thebibliography}

\end{document}